\documentclass[usenatbib,letter]{mnras}

\usepackage{graphicx}
\usepackage{amsmath}
\usepackage{amssymb}

\newcommand{\p}{\partial}

\title[Diffusion in superconducting neutron stars]{Diffusion as a leading dissipative mechanism \\ in superconducting neutron stars}

\author[K. Y. Kraav et al.]
{K. Y. Kraav,
M. E. Gusakov,
and E. M. Kantor
\\
Ioffe Institute, Polytekhnicheskaya 26, St.-Petersburg 194021, Russia}

\date{Accepted 2021 July 6. Received 2021 July 5;
in original form 2021 February 17}

\pubyear{2021}

\begin{document}
\label{firstpage}

\pagerange{\pageref{firstpage}--\pageref{lastpage}}

\maketitle

\begin{abstract}

Despite the fact that different particle species can diffuse with respect to each other in neutron star (NS) cores,
the effect of particle diffusion on various phenomena associated with NS oscillations is usually ignored.
Here we demonstrate that the diffusion 
can be extremely powerful dissipative mechanism in superconducting NSs. 
In particular, it can be much more efficient than the shear and bulk viscosities. 
This result has important implications for the damping times of NS oscillations,
development and saturation of dynamical instabilities in NSs, 
and for the excitation and coupling of oscillation modes during the late inspiral of binary NSs.
\end{abstract}


%%%%%%%%%%%%%%%%%%%%%%%%%%%%%%%%%%%%%%%%%%%%%%%%%%%%%%%%%%%%%%%%%% 
\section{Introduction}
%%%%%%%%%%%%%%%%%%%%%%%%%%%%%%%%%%%%%%%%%%%%%%%%%%%%%%%%%%%%%%%%%% 
%
Neutron stars (NSs) are unique astrophysical objects offering 
an opportunity to probe the properties of the superdense matter under the most extreme conditions 
far beyond those reproduced in terrestrial experiments. 
These properties can be constrained, in particular, by confronting various observational 
manifestations of stellar oscillations with the theoretical models \citep{sw09,sm14,ajh14,sio18,maselli20,kgd20}.

An NS can start oscillating
either as a result of an internal instability
or external perturbation.
Whether an oscillation mode
can actually be excited 
depends on the interplay between the excitation rate 
and on how efficiently dissipative mechanisms 
in the stellar matter counteract this excitation. 
Shear and bulk viscosities are the most paid attention to 
dissipative agents, operating in the NS matter \citep{cl87,mhh88}. 
Generally, at temperatures $T\lesssim 5\times 10^8\,\rm K$, it is the shear viscosity, 
that appears to be the strongest one, 
while the effects due to bulk viscosity are comparatively weak 
and can be ignored 
(\citealt{cls90,gyg05,gg18}; but see \citealt{oghf19}).

Although the chemical composition of the stellar matter 
is rather complex and includes different particle species, such as neutrons (n), protons (p), electrons (e), and muons ($\mu$) in the simplest case, 
and also such ``exotic'' particles 
as hyperons and quarks in the more sophisticated models, 
the effect of particle diffusion on NS 
oscillations,
as far as we know, has not yet been investigated.
In this Letter we take a first look into the role of the diffusion
as a dissipative agent that can 
damp 
NS oscillations (in analogy to how the diffusion 
dissipates
the stellar magnetic field, see \citealt{gr92}).
Below we restrict ourselves to the case of 
npe/npe$\mu$ matter and ignore exotica.
It is believed that neutrons and protons in NS interiors can be 
in the superfluid state \citep[see][]{ls01,plps13}.
According to microscopic calculations \citep{gps14, drddwcp16, sc19},
the maximum critical temperature $T_\text{cn}$ of neutron 
superfluidity onset is substantially lower than the proton one, $T_\text{cp}$.
Therefore, for simplicity, neutrons in this initial study 
are treated as normal,
while for protons we consider two possibilities, 
assuming that they are either  normal (normal matter) or
completely superfluid (superconducting matter; all protons form Cooper pairs).
As we demonstrate, 
in superconducting matter diffusion becomes 
the leading dissipative mechanism that strongly accelerates  the
dissipation of various oscillation modes 
and thus makes questionable our current vision of a number of aspects of NS life.
We expect that our basic conclusions will remain unaffected for not too cold NSs even in the presence of neutron superfluidity, $T\lesssim T_\text{cn}$ (see a comment in the end of Sec.\ 4).

%%%%%%%%%%%%%%%%%%%%%%%%%%%%%%%%%%%%%%%%%%%%%%%%%%%%%%%%%%%%%%%%%% 
\section{The effect of diffusion}
%%%%%%%%%%%%%%%%%%%%%%%%%%%%%%%%%%%%%%%%%%%%%%%%%%%%%%%%%%%%%%%%%% 
%
Let us first discuss the effect of diffusion 
on the damping of sound waves
perturbed in 
the initially homogeneous unmagnetized
npe-matter. 
Each particle species $\alpha$ ($\alpha=$n, p or e) 
is characterized by the electric charge $e_{\alpha}$, 
number density $n_{\alpha}$, 
relativistic chemical potential $\mu_{\alpha}$ (including the rest mass),
and the velocity $\pmb{v}_{\alpha}$.
In superconducting matter $\pmb{v}_\text{p}$ is the velocity
of proton Bogoliubov thermal excitations.
Proton superconductivity adds one more velocity field to the system -- 
the proton superfluid velocity $\pmb{v}_\text{sp}$.
Thanks to the electromagnetic interaction protons and electrons move together,
so that, to a very high precision, their number densities and currents coincide \citep[see][]{braginskii65,mendell91,gd16,dgs20}
\begin{equation}
n_\text{e}=n_\text{p}, \quad \quad n_\text{e} {\pmb v}_\text{e}=n_\text{p ex} {\pmb v}_\text{p}+(n_\text{p}-n_\text{p ex}) {\pmb v}_\text{sp},
\label{quasi}
\end{equation}
where $n_\text{p ex}$ is the number density of proton thermal excitations;
in normal matter $n_\text{p ex}=n_\text{p}$.
An unperturbed matter is supposed to be in $\beta$-equilibrium, $\delta \mu_0\equiv \mu_{\text{n}0}-\mu_{\text{p}0}-\mu_{\text{e}0}=0$ (hereafter index `0' refers to the unperturbed quantities). 
Beta-processes are too slow \citep{ykgh01} to affect oscillations in sufficiently cold matter
and will be neglected in what follows.

Within the framework developed in \cite{braginskii65}, 
the friction force between the species $\alpha$ and $\beta$ is 
$J_{\alpha\beta} \pmb{w}_{\alpha\beta}$, 
where $\pmb{w}_{\alpha\beta} \equiv \pmb{v}_{\alpha}-\pmb{v}_{\beta}$
and $J_{\alpha\beta}$ is the momentum transfer rate. 
The energy leakage due to diffusion equals the work done by the friction force per unit time 
and can be written as an integral over the stellar volume $V$ \citep[see][]{braginskii65,gko17} 
\begin{gather}
\dot{E}_\text{diff}=-\frac{1}{2}\int \sum_{\alpha\beta} J_{\alpha\beta} w_{\alpha\beta}^2 ~\text{d}V. \label{Ediff}
\end{gather}
Although \cite{braginskii65} discussed Eq.\ (\ref{Ediff}) 
in application to the
normal matter, 
the same expression,
 in principle,
is
valid in the superfluid/superconducting matter
provided that the proton velocity in Eq.\ (\ref{Ediff})
is that of the proton thermal excitations
(because only thermal excitations can scatter from other particle species; see, e.g., \citealt{aggk81}).

Relative velocities in Eq.\ (\ref{Ediff}) 
depend on the efficiency of particle collisions 
(more effective collisions correspond 
to smaller relative velocities) 
and  should be expressed 
through other variables
describing perturbation. 
When particle collisions are much more frequent than 
the oscillation frequency $\omega$,
the system is in the hydrodynamic regime.
In this regime the linearized Euler-like equations governing 
perturbations in the {\it normal} matter take the form 
\citep{braginskii65,gr92,dgs20} ($\alpha=$n,p,e)
\begin{eqnarray}
\label{Brag_eqs_red}
\frac{n_{\alpha 0}\mu_{\alpha 0}}{c^2}\frac{\p\pmb{v}_\alpha}{\p t}=e_{\alpha} n_{\alpha 0}\pmb{E}-n_{\alpha 0}\pmb{\nabla}\mu_{\alpha}-\sum_{\beta}J_{\alpha\beta}\pmb{w}_{\alpha\beta},
\end{eqnarray}
where $\pmb{E}$ is the electric field 
and $c$ the speed of light. 
The hydrodynamic limit corresponds 
to large values of $J_{\alpha\beta}$ 
such that ${w}_{\alpha\beta}$ is much smaller 
than a typical hydrodynamic velocity $v$. 
Thus, in what follows ${w}_{\alpha\beta}$ can be set to zero 
{\it unless} it is multiplied by $J_{\alpha\beta}$. 

Eqs.\ (\ref{Brag_eqs_red}) are widely known 
in the neutron-star and plasma literature (e.g., \citealt{gko17} and references therein)
and can be derived from the Boltzmann transport equations for each particle species $\alpha$ in a manner,
completely analogous to the derivation in \cite{ys91}.
Note that, in these equations only diffusion dissipation is allowed for, while other
dissipative mechanisms (e.g., viscosity) are ignored.
This is justified in the hydrodynamic regime, in which effects of different 
dissipation mechanisms can be studied separately (e.g., \citealt{ll87}).

Summing up Eq.\ \eqref{Brag_eqs_red} divided by $n_{\alpha 0}$ for protons and electrons and subtracting Eq.\ \eqref{Brag_eqs_red} for neutrons divided by $n_{\text{n}0}$, we, using Eq.\ (\ref{quasi})
and the $\beta$-equilibrium condition $\delta \mu_0=0$,
arrive at the equation
\begin{gather}
\label{eq4}
-\frac{\mu_{\text{n}0}}{c^2}\frac{\p\pmb{w}_\text{np}}{\p t}=\pmb{\nabla}\delta \mu+\left[\frac{J_\text{np}}{n_{\text{p} 0}}+\frac{J_\text{en}}{n_{\text{e}0}}+\frac{J_\text{np}+J_\text{en}}{n_{\text{n}0}}\right]\pmb{w}_\text{np}.
\end{gather}
The inertial term in the left-hand side of this equation, which can be estimated as
$-\mu_{{\rm n}0} \omega{\pmb w}_{\rm np}/c^2$,
is much smaller than the last term in the right-hand side
if $\mu_{{\rm n}0}\omega/c^2 \ll J_{ni}/n_{i}$ ($i=n$, $p$, $e$).
Taking into account that $J_{ni}\sim (10^{24}-10^{30}) (T/10^8~{\rm K})^2$~g~cm$^{-3}$~s$^{-1}$
(see, e.g., Section VII in \citealt{dgs20}), 
this inequality is always satisfied for typical neutron-star conditions,
so that we can neglect the inertial term in (\ref{eq4}).
This allows us to
express the relative velocities through $\delta \mu$ 
and transform the energy loss rate due to diffusion (\ref{Ediff}) into
\begin{eqnarray}
\dot{E}_\text{diff}
\approx-\int\frac{1}{J_\text{np}}\biggl[\frac{n_{\text{n}0}n_{\text{e}0}}{n_{\text{b}0}}\pmb{\nabla}\delta\mu\biggr]^2 \text{d}V,
\label{Edot_diff}
\end{eqnarray}
where $n_\text{b}=n_\text{n}+n_\text{p}$ is the baryon number density. 
To derive (\ref{Edot_diff})
we neglected $J_{\rm en}$ compared to $J_{\rm np}$ 
since in normal matter collisions between protons and neutrons are extremely efficient 
due to strong interactions and $J_{\rm np}/J_{\rm en} \sim 10^5$ \citep[see][]{dgs20}.

In what follows, we  
consider the case of strong proton superconductivity, 
$T \ll T_{\rm cp}$, suitable for not too young NSs.
At $T \ll T_{\rm cp}$ $n_{\rm p\, ex} \rightarrow 0$ and Eq.\ (\ref{Brag_eqs_red}) for protons
should be replaced with 
the Josephson equation for the superconducting proton component \citep{putterman74}
\begin{eqnarray}
\label{sf_eq}
\frac{\mu_{\text{p} 0}}{c^2}\frac{\p\pmb{v}_\text{sp}}{\p t}=e_\text{p} \pmb{E}-\pmb{\nabla}\mu_\text{p}.
\end{eqnarray}
Moreover, in this limit 
superconductivity strongly suppresses 
scattering processes involving protons 
(the number of proton thermal Bogoliubov excitations 
is exponentially suppressed),
so that the proton-related momentum transfer rates tend to zero, $J_{\text{p}\alpha}=0$.
In these conditions Eq.\ (\ref{Ediff}) reduces to
\begin{eqnarray}
\label{Edot_diff_sf1}
\dot{E}_\text{diff}=-\int\frac{1}{J_\text{en}}\biggl[\frac{n_{\text{n}0}n_{\text{e}0}}{n_{\text{b}0}}\pmb{\nabla}\delta\mu\biggr]^2 \text{d}V.
\end{eqnarray}
To derive (\ref{Edot_diff_sf1}) we make use of Eq.\ (\ref{quasi}) with $n_{\rm p\, ex}=0$, 
Eq.\ (\ref{Brag_eqs_red}) for neutrons and electrons, 
Eq.\ (\ref{sf_eq}), and the equality $\delta \mu_0=0$.

When dissipation is weak, the imbalance
$\delta \mu$ in Eqs.~(\ref{Edot_diff}) and (\ref{Edot_diff_sf1}) 
can be calculated using the equations of nondissipative hydrodynamics 
(see, e.g., Section 3.1), 
which are the same
in normal and superconducting matter. 
Then Eqs.~(\ref{Edot_diff}) and (\ref{Edot_diff_sf1}) imply that dissipation 
in superconducting matter
is by a factor of 
$\sim J_\text{np}/J_\text{en}\sim 10^5$ more efficient than in normal matter.

It can be shown that Eqs.\ (\ref{Edot_diff})
and (\ref{Edot_diff_sf1}) are equally applicable to the 
inhomogeneous npe-matter of Newtonian stars.
Similar equations valid in General Relativity (GR)
can 
be derived within the framework of
relativistic multi-fluid dissipative hydrodynamics
developed 
in \cite{dgs20,dg21}.

%%%%%%%%%%%%%%%%%%%%%%%%%%%%%%%%%%%%%%%%%%%%%%%%%%%%%%%%%%%%%%%%%% 
\section{Oscillation damping times}
%%%%%%%%%%%%%%%%%%%%%%%%%%%%%%%%%%%%%%%%%%%%%%%%%%%%%%%%%%%%%%%%%% 

To estimate the effect of diffusion on damping of sound waves, as well
as on damping of global 
f-, p-, g-, and r-modes,
we 
confront here
the damping (e-folding) times due to diffusion, 
$\tau_{\rm diff}=-2E/\overline{\dot{E}}_{\rm diff}$, 
with those due to shear viscosity, 
$\tau_\eta=-2E/\overline{\dot{E}}_\eta$, 
where   
$\overline{\dot{E}}_\eta$ and $\overline{\dot{E}}_{\rm diff}$ are the corresponding dissipation rates averaged over the oscillation period and $E$ is the oscillation energy \citep{ll87}.

In all numerical calculations below we employ the BSk24 equation of state
\citep{gcp13}, 
allowing for muons, and adopt shear viscosity coefficients 
and momentum transfer rates from \cite{shternin18} and \cite{dgs20}. 
In the case of superconducting matter we use 
equation (28) of \cite{shternin18}
for the shear viscosity 
and neglect the effect of proton superconductivity on $J_\text{en}$.

%%%%%%%%%%%%%%%%%%%%%%%%%%%%%%%%%%%%%%%%%%%%%%%%%%%%%%%%%%%%%%%%%% 
\subsection{Sound waves}
%%%%%%%%%%%%%%%%%%%%%%%%%%%%%%%%%%%%%%%%%%%%%%%%%%%%%%%%%%%%%%%%%% 

To calculate $\tau_{\rm diff}$ for sound waves, 
we expand 
$\delta \mu$ in Eqs.~(\ref{Edot_diff}) and (\ref{Edot_diff_sf1}) 
[which is, generally, a function of all particle number densitities, 
$\delta \mu=\delta\mu(n_{\rm n},n_{\rm p},n_{\rm e})$]
as
\begin{align}
	\delta \mu = \sum_\alpha 
	   \frac{\partial \delta \mu}{\partial n_\alpha} 
	   \delta n_\alpha = 
	\sum_\alpha 
	\frac{\partial \delta \mu}{\partial n_\alpha}
	 \frac{n_{\alpha 0} v k}{\omega},
	\label{dmu}
\end{align}
where $k$ is the wave number, and we used the  
continuity equations,
$-\omega \delta n_\alpha+k n_{\alpha 0}v=0$,
to express the particle number density perturbations $\delta n_\alpha$ ($\alpha=\rm n,p,e$) 
through the fluid velocity $v=v_0 \, {\rm cos}(k x -\omega t)$ 
(which is the same for all particle species in the nondissipative limit).
For sound waves, the oscillation energy per unit volume is 
$E=\sum_\alpha \mu_{\alpha0} n_{\alpha0} v_0^2/(2c^2)$, 
thus
\begin{align}
\tau_{\rm diff}=\mathcal{J}\frac{2\omega^2}{k^4}\left(\frac{n_{\rm b0}}{n_{\rm n0}n_{\rm e0}}\right)^2 \left(\sum_\alpha \frac{\partial \delta \mu}{\partial n_\alpha}n_{\rm \alpha 0}\right)^{-2}\sum_\alpha 
\frac{\mu_{\alpha 0} n_{\alpha0}}{c^2},
\label{tdiff}
\end{align}
where $\mathcal{J}$ stands for $J_{\rm np}$ in normal matter and for $J_{\rm en}$ in superconducting matter.

\begin{figure}
\center{\includegraphics[width=0.77\linewidth]{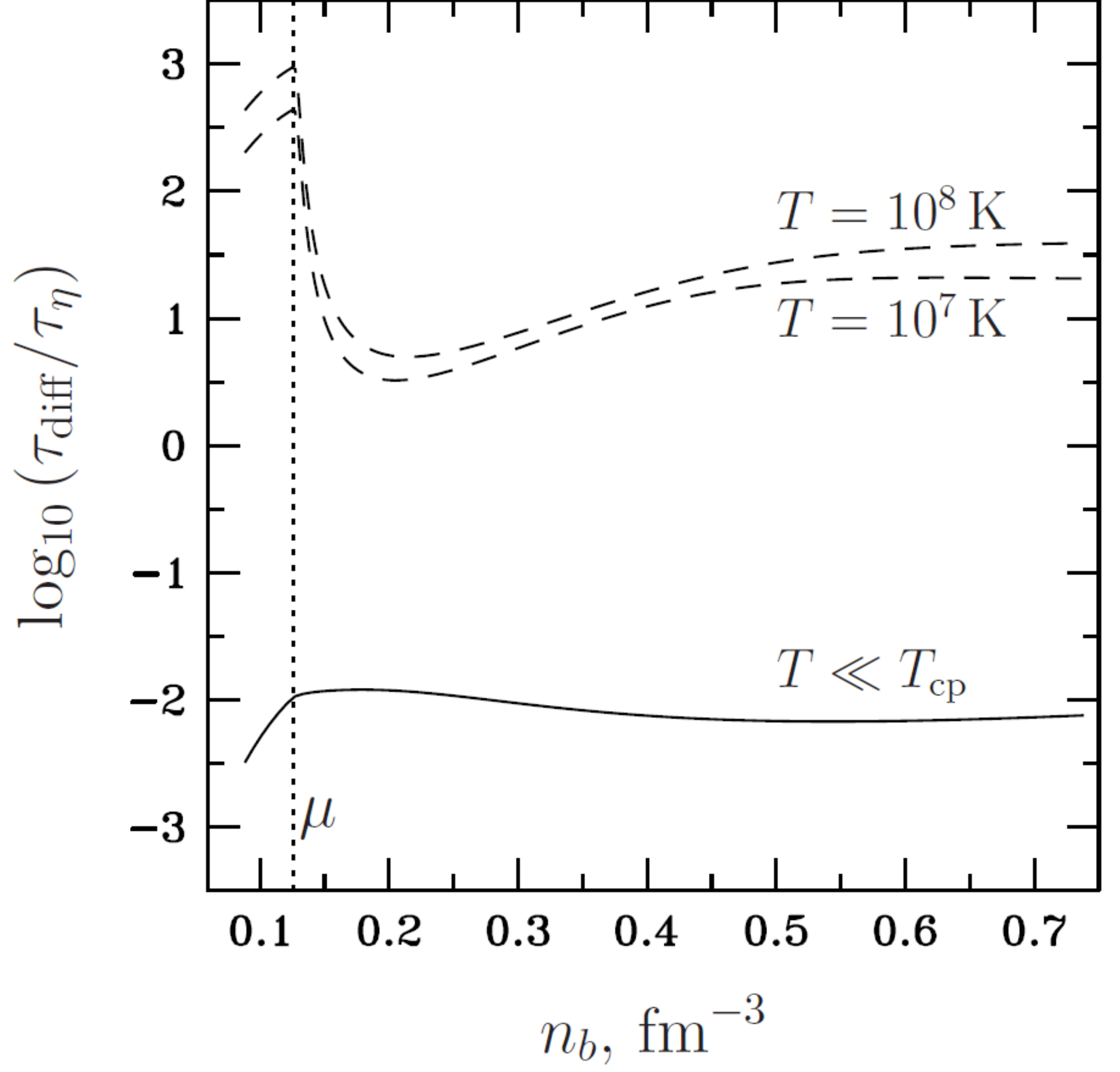}}
\caption{$\tau_{\rm diff}/\tau_\eta$ versus $n_\text{b}$ for sound waves. 
Vertical dots denote the muon onset density.}
\label{sound}
\end{figure}

In order to proceed to higher densities, 
where muons are present, 
we generalize the results derived above 
to the case of npe$\mu$ matter. 
The ratio $\tau_{\rm diff}/\tau_\eta$ 
is plotted in Fig.\ \ref{sound} as a function of $n_\text{b}$
for normal (dashes) and strongly superconducting ($T\ll T_\text{cp}$, solid line) matter. 
In superconducting matter $\tau_{\rm diff}$ and $\tau_\eta$ 
obey almost the same temperature dependence ($\propto T^2$)
and their ratio is almost temperature-independent 
($\tau_{\rm diff}$ is approximately $\propto T^2$ because $J_{\rm en}$ and $J_{\rm \mu n}$ are $\propto T^2$, see \citealt{dgs20}; 
small deviation from this scaling is caused by the coefficient $J_{\text{e}\mu}$, which has a more complicated behavior, but whose contribution is small).
This is not the case in normal matter for which we present 
$\tau_{\rm diff}/\tau_\eta$ for 
$T=10^7$ and $10^8$~K.
The curves do not vary with 
$k$ or $\omega$, 
since both damping times scale as $k^{-2}\propto\omega^{-2}$ 
(see \citealt{ll87}
and note that $\omega\propto k$ for sound waves).
In normal npe-matter particles are locked to each other: 
neutrons are locked to protons due to frequent collisions caused by strong interaction, 
while electrons are locked to protons by electromagnetic interaction. 
As a result, diffusion is inefficient. 
Appearance of muons allows charged particles to move with respect to each other, 
and the efficiency of diffusion increases strongly. 
However, our results imply that, anyway, 
diffusion is less effective in normal matter in the whole range of densities than the shear viscosity.
At the same time, in superconducting matter neutrons are free to move with respect to protons 
and it is the diffusion, that becomes the dominant channel of energy losses.
As we demonstrate below, this conclusion remains correct 
also for global oscillatory modes of NSs.

%%%%%%%%%%%%%%%%%%%%%%%%%%%%%%%%%%%%%%%%%%%%%%%%%%%%%%%%%%%%%%%%%% 
%%%%%%%%%%%%%%%%%%%%%%%%%%%%%%%%%%%%%%%%%%%%%%%%%%%%%%%%%%%%%%%%%% 

%%%%%%%%%%%%%%%%%%%%%%%%%%%%%%%%%%%%%%%%%%%%%%%%%%%%%%%%%%%%%%%%%%
\subsection{Global oscillation modes}
%%%%%%%%%%%%%%%%%%%%%%%%%%%%%%%%%%%%%%%%%%%%%%%%%%%%%%%%%%%%%%%%%%

Using the formalism developed in \cite{dgs20} and \cite{dg21}, 
we calculate the damping times 
of relativistic f-, p-, and g-modes, and Newtonian r-modes for an NS 
in the Cowling approximation (e.g., \citealt{ls90}). 
We consider a three-layer NS consisting of the barotropic crust, 
npe outer core, and npe$\mu$ inner core. 
We assume that protons are strongly superconducting, 
$T_\text{cp}\gg T$, while neutrons are normal.
The oscillation eigenfunctions and eigenfrequencies in the absence of dissipation 
(in particular, the imbalance $\delta \mu$)
are calculated with our codes developed in \cite{kg14,kg17}.

\begin{figure}
\center{\includegraphics[width=0.77\linewidth]{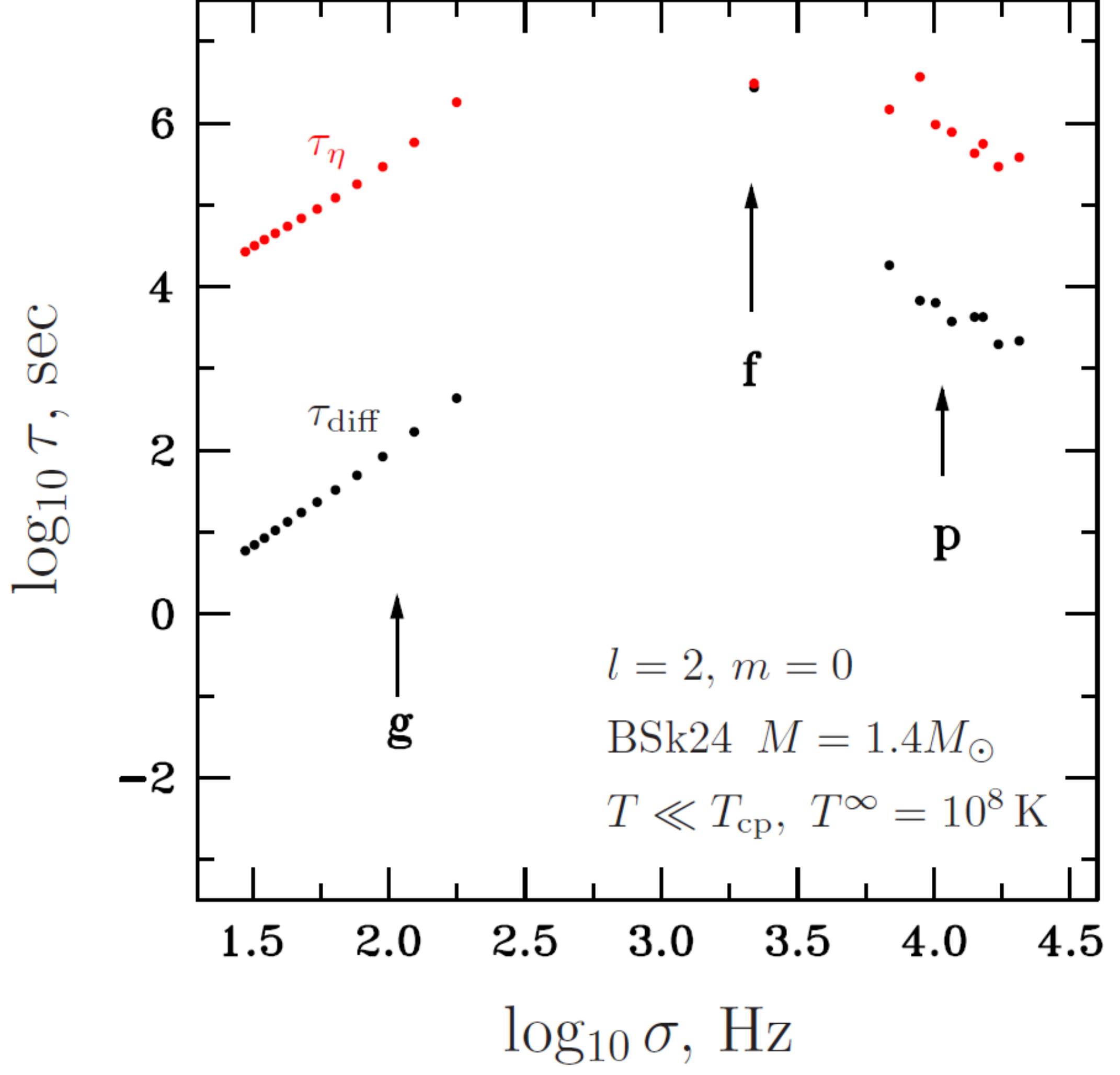}}
\caption{Damping times for the first ($l=2,\,m=0$) f-, p-, and g-modes in superconducting NS due to shear viscosity (upper red points) and diffusion (lower black points) versus the mode eigenfrequency.}
\label{modes}
\end{figure}

%%%%%%%%%%%%%%%%%%%%%%%%%%%%%%%%%%%
{\bf f-, p-, and g-modes: }
%%%%%%%%%%%%%%%%%%%%%%%%%%%%%%%%%%%
%
Damping times
versus eigenfrequency $\sigma\equiv \omega/(2\pi)$
are shown in Fig.\ \ref{modes} 
for the first ($l=2$, $m=0$) eigenmodes 
of an NS with the mass $M=1.4M_\odot$ 
and redshifted
internal stellar temperature $T^\infty=10^8\,\rm K$
(as seen by a distant observer). 
Upper points represent dissipation due to shear viscosity, 
$\tau_{\eta}$, while lower points show diffusion damping times, $\tau_{\rm diff}$. 

%%%%%%%%%%%%%%%%%%%%%%%%%%%%%%%%%%%%%%%%%%%%%%%%%%%%%%%%%%%%%%%%%% 
%%%%%%%%%%%%%%%%%%%%%%%%%%%%%%%%%%%%%%%%%%%%%%%%%%%%%%%%%%%%%%%%%% 

One can see that for p-modes $\tau_{\eta}$ 
exceeds $\tau_{\rm diff}$ by approximately two orders of magnitude. 
The same result was obtained for sound waves (Fig.\ \ref{sound}), 
which is not surprising since p-modes are their close relatives. 
In analogy to sound waves, both damping times fall simultaneously 
with increasing number of nodes in a p-mode, 
or, equivalently, 
with increasing $\sigma$: $\tau_{\eta}\propto \tau_{\rm diff}\propto \sigma^{-2}$.

For g-modes the difference between $\tau_{\eta}$ 
and $\tau_{\rm diff}$ is even larger --- almost four orders of magnitude.
For these low-frequency oscillations $\sigma\propto L$, 
where $L$ is the lengthscale of the perturbation, 
while higher harmonics are dominated 
by the $\theta$-component of the velocity $v_\theta \sim v_r\sigma_0/\sigma$, 
where $v_r$ is the radial velocity and $\sigma_0$ is the eigenfrequency of the main harmonic. 
Moreover, the Eulerian perturbation of the pressure is very small, 
which leads to $\nabla \delta \mu \propto v_r/(\sigma L)$. 
As a result, $\tau_{\eta}\propto \tau_{\rm diff}\propto \sigma^2$ for g-modes.
Note that 
in case of g-modes
diffusion remains the most powerful dissipation mechanism even for normal NSs,
e.g.  $\tau_{\eta}/\tau_{\rm diff}\approx 4.8$ for the main g-mode harmonic.
In turn, 
the efficiency of diffusion for the f-mode 
is strongly suppressed 
($\tau_{\eta}$ and $\tau_{\rm diff}$ almost coincide) 
since this mode is almost incompressible and chemical potential imbalances 
are practically not perturbed in the course of oscillations.

%%%%%%%%%%%%%%%%%%%%%%%%%%%%%%%%%%%
{\bf r-modes: }
%%%%%%%%%%%%%%%%%%%%%%%%%%%%%%%%%%%
%
The rotational, predominantly toroidal modes -- r-modes --
puzzle NS community since 1998, 
when they were predicted to be unstable 
due to gravitational radiation \citep{andersson98,fm98}.
Confronting the r-mode excitation and dissipation timescales 
defines
the 
instability window, that is the region 
on the 
$\nu-T^\infty$ plane,
where dissipation cannot counteract the r-mode growth \citep{haskell15} 
($\nu$ is the NS rotation frequency). 
Despite predictions of the modeling
that an NS cannot stay in the instability window for
any considerable amount of time \citep[see][]{levin99}, 
numerous sources have been observed there \citep{gck14a}.
Revealing
some strong dissipative mechanism, that 
has not yet been
identified
could reconcile theory and observations. 
Here we examine whether diffusion could serve as such a mechanism.

\begin{figure}
\center{\includegraphics[width=0.77\linewidth]{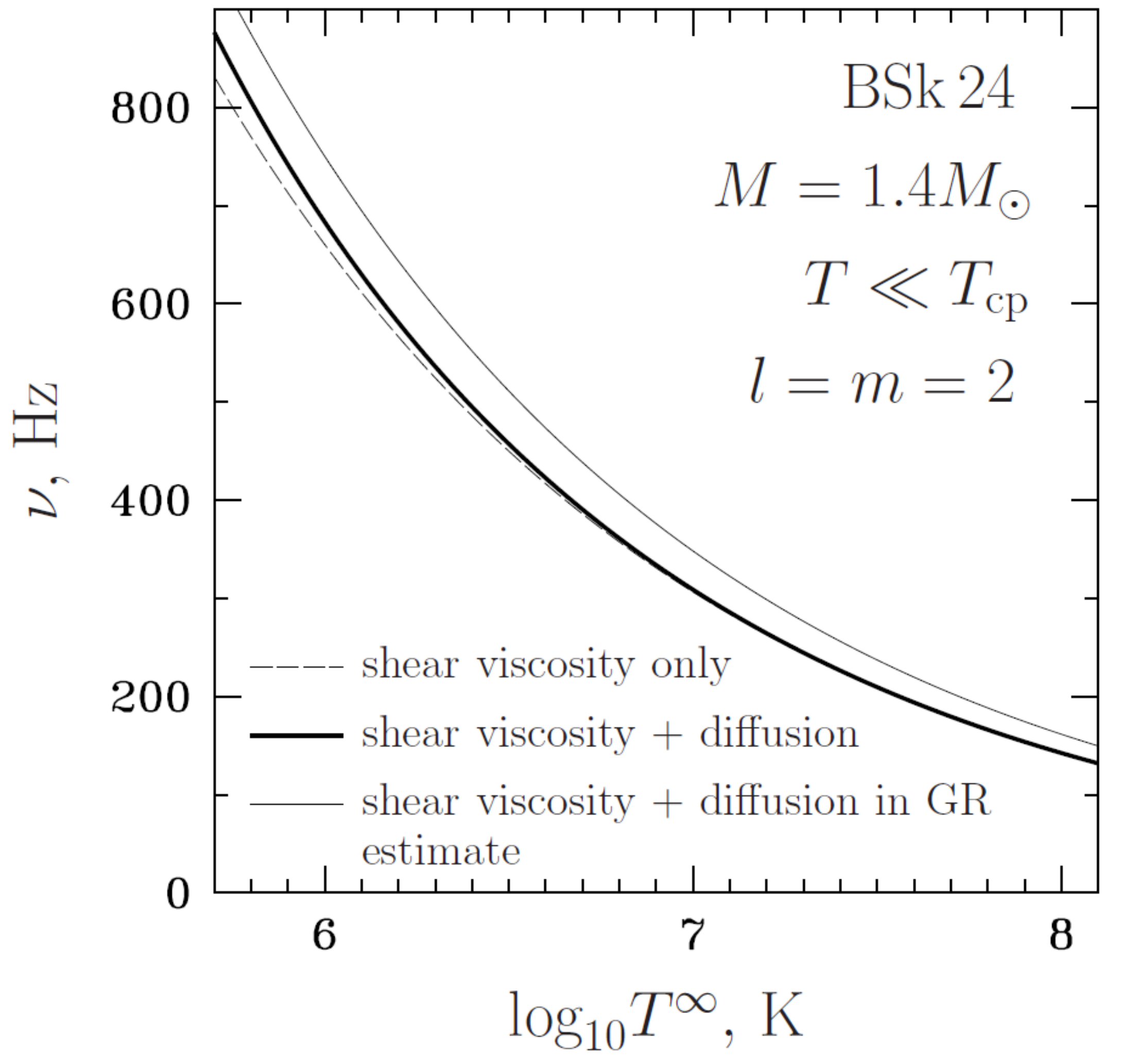}}
\caption{Instability curves for $l=m=2$ r-mode in superconducting (but nonsuperfluid) NS. We extend the temperature range to extremely low values 
to
illustrate the behaviour of the curves at large $\nu$.}
\label{r-modes}
\end{figure}

We calculate damping times for the most unstable $l=m=2$ r-mode,
treating perturbations within the Newtonian framework in the Cowling approximation,
and adopting the same stellar model and microphysics input as above. 
In our calculations we assume that 
$\nu$ is small compared to the Kepler frequency, $\nu_{\rm K}$, 
and expand all the perturbations 
in the parameter $\nu/\nu_{\rm K}$. 
Fig. \ref{r-modes} shows the resulting instability curves (boundaries of the instability window), 
where the r-mode excitation is balanced by
the shear viscosity only (dashes) and by the combined action of 
shear viscosity and diffusion (thick solid line). 
Above the curves the r-mode is unstable.

%%%%%%%%%%%%%%%%%%%%%%%%%%%%%%%%%%%%%%%%%%%%%%%%%%%%%%%%%%%%%%%%%% 
%%%%%%%%%%%%%%%%%%%%%%%%%%%%%%%%%%%%%%%%%%%%%%%%%%%%%%%%%%%%%%%%%% 

Note that for the r-mode $\sigma\propto \nu$, 
while perturbations of thermodynamic quantities, 
in particular chemical potentials, are suppressed by a factor of $\nu^2/\nu_{\rm K}^2$. 
As a result, $1/\tau_{\rm diff}$ turns out to be $\propto \nu^2$ 
[see Eq.\ (\ref{Edot_diff_sf1})], 
while $1/\tau_\eta$ does not depend on $\nu$. 
Thus, while at high $\nu$ diffusion is as effective as shear viscosity%
%
%%%%%%
\footnote{Similarly to f-mode, the r-mode is almost incompressible \citep{lmo99}
and the effect of diffusion is not so pronounced.}, 
%%%%%%
%
at lower $\nu$ diffusion is negligible.
Consequently, in the Newtonian framework 
the effect of diffusion on the instability curve 
is smaller at higher $T^\infty$, 
since the curve 
$\nu(T^\infty)$ in Fig.\ \ref{r-modes} 
is a decreasing function of temperature.

Notice, however, that in GR perturbations 
of chemical potentials are $\propto\nu/\nu_{\rm K}$ 
(see equation 17 in \citealt{lfa03} and also \citealt{laf01}),
hence $1/\tau_{\rm diff}$ does not scale with $\nu$.
We have not carried out r-mode calculations in full GR yet. 
However, to get an impression of how efficient diffusion in GR may be, 
we assumed that Newtonian approach and GR 
give similar results for $\tau_{\rm diff}$ at $\nu=\nu_{\rm K}$ 
(since chemical potential perturbations are not suppressed at $\nu=\nu_{\rm K}$).
Note that, while the expansion parameter 
$\nu/\nu_{\rm K}$ is not small at $\nu=\nu_{\rm K}$, 
we can still formally solve the expanded oscillation equations 
at $\nu=\nu_{\rm K}$ and rescale the result to smaller $\nu$.
Then we calculate $\tau_{\rm diff}$ at $\nu=\nu_{\rm K}$
in the Newtonian framework and assume that in GR $\tau_{\rm diff}$ equals this value at any $\nu$. 
The resulting GR instability curve due to diffusion and shear viscosity is shown by thin solid line
and noticeably differs from the thick one.
We emphasize that this is only an estimate, 
an accurate calculation will be published elsewhere.

%%%%%%%%%%%%%%%%%%%%%%%%%%%%%%%%%%%%%%%%%%%%%%%%%%%%%%%%%%%%%%%%%% 
\section{Conclusions}
%%%%%%%%%%%%%%%%%%%%%%%%%%%%%%%%%%%%%%%%%%%%%%%%%%%%%%%%
%
We propose that  
particle diffusion 
can be a very efficient dissipative mechanism in NSs.
In contrast to the mutual friction dissipation (e.g., \citealt{haskell15}), 
this mechanism does not need superfluid vortices in order to operate.
We compare damping of sound waves, as well as of f-, p-, g-, and r- modes 
due to diffusion and shear viscosity 
in NSs composed of neutrons, protons, and leptons. 
We find that, when protons are normal, 
the effect of diffusion on stellar oscillations is relatively small 
and can be ignored for all modes except for g-modes. 
In contrast, 
for superconducting protons
diffusion leads to the very fast damping of oscillations, 
especially in the case of sound waves, p- and g- modes, 
leaving shear viscosity (which is believed to be the key dissipative mechanism) far behind.

Our results imply that damping times of all oscillation modes should be revisited. 
Every physical phenomenon, 
dealing with the development of one or another hydrodynamic instability 
--- for example, resonant excitation of modes \citep{yw17}
and $p$-$g$ tidal instability \citep{weinberg16,ah18,pg19}
during NS-NS and NS-BH binary inspirals;
excitation of r-modes in rapidly rotating NSs \citep{haskell15}; 
saturation of r-modes due to non-linear coupling 
to the inertial-gravity modes \citep{btw07};
instability onset due to precession \citep{gaj09}; 
glitch models based upon the excitation of 
hydrodynamic instabilities in the core (e.g., \citealt{ga09}) 
and others --- has to be reconsidered in order to check whether 
these instabilities can survive in the presence of the discussed here powerful diffusion dissipation. 
Further, strong 
dissipation
due to diffusion 
may reduce the excitation time of the viscosity-driven secular instability in NSs \citep{lindblom87,sl96,andersson03,shapiro04}.
We therefore conclude that diffusion may have an important effect on the interpretation of gravitational signal produced in binary NS late inspirals, glitches, as well as on the interpretation of observational properties of rapidly rotating NSs.

Finally, let us briefly discuss how the magnetic field and possible neutron superfluidity may affect our results.
If protons form a type-II superconductor, 
then the magnetic flux is confined to flux tubes.
Particles may scatter off the magnetic field of the tubes
or off the 
proton excitations localized inside the tubes.
Both these mechanisms tend to reduce $w_{\alpha\beta}$ and hence decrease dissipation.
However, our estimates show that the relative particle velocities {\it along}
the magnetic field lines
will not be substantially reduced 
for not too high magnetic fields $B\ll H_\text{c2}$,
where $H_\text{c2}\sim 10^{15}-10^{16}$~G is the second critical magnetic field.
Since the velocity field of a generic oscillation mode has a component along the magnetic field lines, diffusion at $B\ll H_\text{c2}$
should remain almost as effective as in the absence of the magnetic field.
Similar conclusion also applies to the situation when protons form 
type-I superconductor, which is likely the case in the inner NS cores. 
As for neutron superfluidity, it
could increase diffusive dissipation, 
since it suppresses $J_{\text{n}\alpha}$
(i.e., suppresses neutron scattering from other particles).
However, this effect can be compensated by the reduction in the number of ``normal'' 
neutrons participating in the scattering processes. 
Moreover, neutron superfluidity modifies the hydrodynamic flows, 
which can also affect diffusion. 
A combined study of all these 
effects is important and deserves a special consideration.

\section*{Acknowledgements}
MEG was partly supported by RFBR [Grant No. 19-52-12013].

\section*{Data availability}
The data underlying this article are available in the article.

\label{lastpage}

\end{document}